%% file: taslp25-submission.tex
\PassOptionsToPackage{pdftex,final,breaklinks,hidelinks,colorlinks=false,bookmarks=false,bookmarksnumbered,plainpages=false,pdfpagelabels,pdfborder={0 0 0},hyperindex}{hyperref}
\PassOptionsToPackage{table,svgnames, x11names, dvipsnames}{xcolor}
\PassOptionsToPackage{scaled=.8}{beramono}
\documentclass[letterpaper,journal]{IEEEtran}

\usepackage[utf8]{inputenc}
\usepackage[T1]{fontenc}
\usepackage[scaled=.87]{beramono} 

\usepackage[noteubner, frozen]{showcode}
\usepackage[greek,english]{babel}
\usepackage{url}
\usepackage{soul}
\usepackage{pifont}
\newcommand{\dingred}[1]{{\color{BrickRed}\ding{#1}}}
\newcommand{\dinggreen}[1]{{\color{ForestGreen}\ding{#1}}}
\newcommand{\dingblue}[1]{{\color{Blue}\ding{#1}}}

\definecolor{MyRawSienna}{RGB}{173, 52, 50}
\definecolor{MyOrange}{RGB}{254, 113, 28}
\definecolor{MyBloodRed}{RGB}{153, 0, 100}
\definecolor{MyTealBlue}{RGB}{1, 172, 175}
\definecolor{MyRoyalBlue}{RGB}{1, 87, 180}
\definecolor{MyViolet}{RGB}{61, 60, 158}
\definecolor{NegativeMyViolet}{RGB}{158, 60, 61}
\usepackage{paralist}

\newcommand{\ourtool}[1]{\texttt{\textbf{#1}}}

\usepackage{tikz}
\usepackage{orcidlink}

\usepackage{amsmath}
\usepackage{mathtools}
\usepackage{centernot}
\usepackage{bm}

\usepackage{float}

\usepackage{caption}

\usepackage[numbers]{natbib}

\title{Sink or SWIM\@: Tackling Real-Time ASR at Scale}
\author{Federico Bruzzone\,\orcidlink{0009-0004-6086-8810}, Walter Cazzola\,\orcidlink{0000-0002-4652-8113}, Matteo Brancaleoni and Dario Pellegrino
   \IEEEcompsocitemizethanks{\IEEEcompsocthanksitem%
        F.~Bruzzone is with the Department of Computer Science,
        Universit\`a degli Studi di Milano, Milan, Italy.
        E-mail: \href{mailto:federico.bruzzone@unimi.it}{federico.bruzzone@unimi.it}.}
    \IEEEcompsocitemizethanks{\IEEEcompsocthanksitem%
        W.~Cazzola (corresponding author) is with the Department of Computer Science,
        Universit\`a degli Studi di Milano, Milan, Italy.\protect\\
        E-mail: \href{mailto:cazzola@di.unimi.it}{cazzola@di.unimi.it}.}
    \IEEEcompsocitemizethanks{\IEEEcompsocthanksitem%
        M.~Brancaleoni is with the Computer Science Division,
        VoiSmart, Milan, Italy.
        E-mail: \href{mailto:matteo.brancaleoni@voismart.it}{matteo.brancaleoni@voismart.it}.}
    \IEEEcompsocitemizethanks{\IEEEcompsocthanksitem%
        D.~Pellegrino is with the Computer Science Division,
        VoiSmart, Milan, Italy.
        E-mail: \href{mailto:dario.pellegrino@voismart.it}{dario.pellegrino@voismart.it}.\vspace*{5pt}}
}

\begin{document}
   \maketitle

   \input{sects/abstract}

   \input{sects/introduction}
   \input{sects/background}
   \input{sects/design}
   \input{sects/evaluation}
   \input{sects/threats}
   \input{sects/conclusion}

\bibliographystyle{IEEEtranN}
\bibliography{local,strings,metrics,foundations,data_structures,programming,software_engineering,software_architecture,dsl,pl,splc,oolanguages,my_work,grammars,security,roles,learning,cop,testing,dsu,distributed_systems,reflection,aosd,pattern,logic,ml+nn}

\vspace*{-1.45cm}
\begin{IEEEbiography}[{\includegraphics[width=1in,height=1.25in,clip,keepaspectratio]{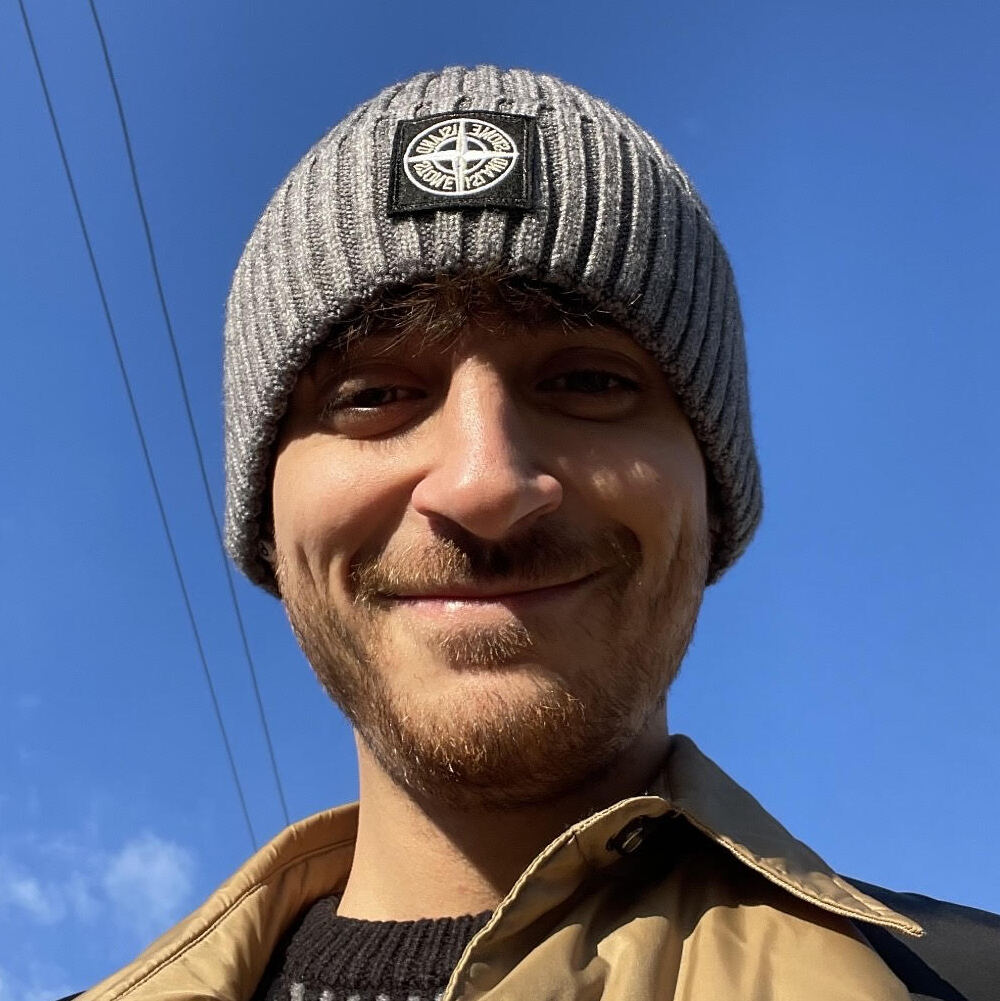}}]{Federico Bruzzone} is currently a Ph.D. student in Computer Science at Universit\`a degli Studi di Milano, Italy. He was born in 2000 and since he was a child he has been passionate about computer science and music. He got his bachelor degree in Musical Computer Science, the master degree in Computer Science and currently he is involved in the research activity of the ADAPT Lab. His main research interests are (but are not limited to) programming languages and compilers, software maintenance and evolution. For any question he can be contacted at \url{federico.bruzzone@unimi.it}
\end{IEEEbiography}
\vspace*{-1.45cm}
\begin{IEEEbiography}[{\includegraphics[width=1in,height=1.25in,clip,keepaspectratio]{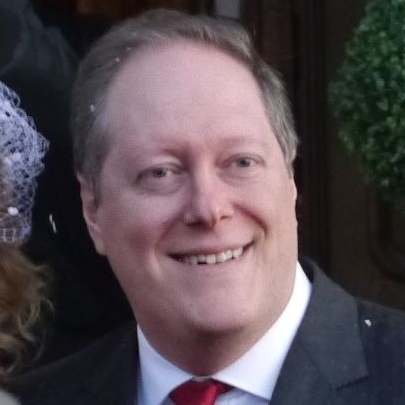}}]{Walter Cazzola} is currently a Full Professor in the Computer Science Department of the Università degli Studi di Milano, Italy and the Chair of the ADAPT laboratory. He designed the mChaRM framework, @Java, [a]C\#, Blueprint programming languages and he is currently involved in the designing and development of the Neverlang language workbench. He also designed the JavAdaptor dynamic software updating framework and its front-end FiGA\@. His research interests include software maintenance, evolution and comprehension, programming methodologies and languages. He served on the program committees or editorial boards of the most important conferences and journals about his research topics. He is associate editor for the Journal of Computer Languages published by Elsevier. More information are available at \url{https://cazzola.di.unimi.it} and he can be contacted at \url{cazzola@di.unimi.it} for any question.
\end{IEEEbiography}
\vspace*{-1.45cm}
\begin{IEEEbiography}[{\includegraphics[width=1in,height=1.25in,clip,keepaspectratio]{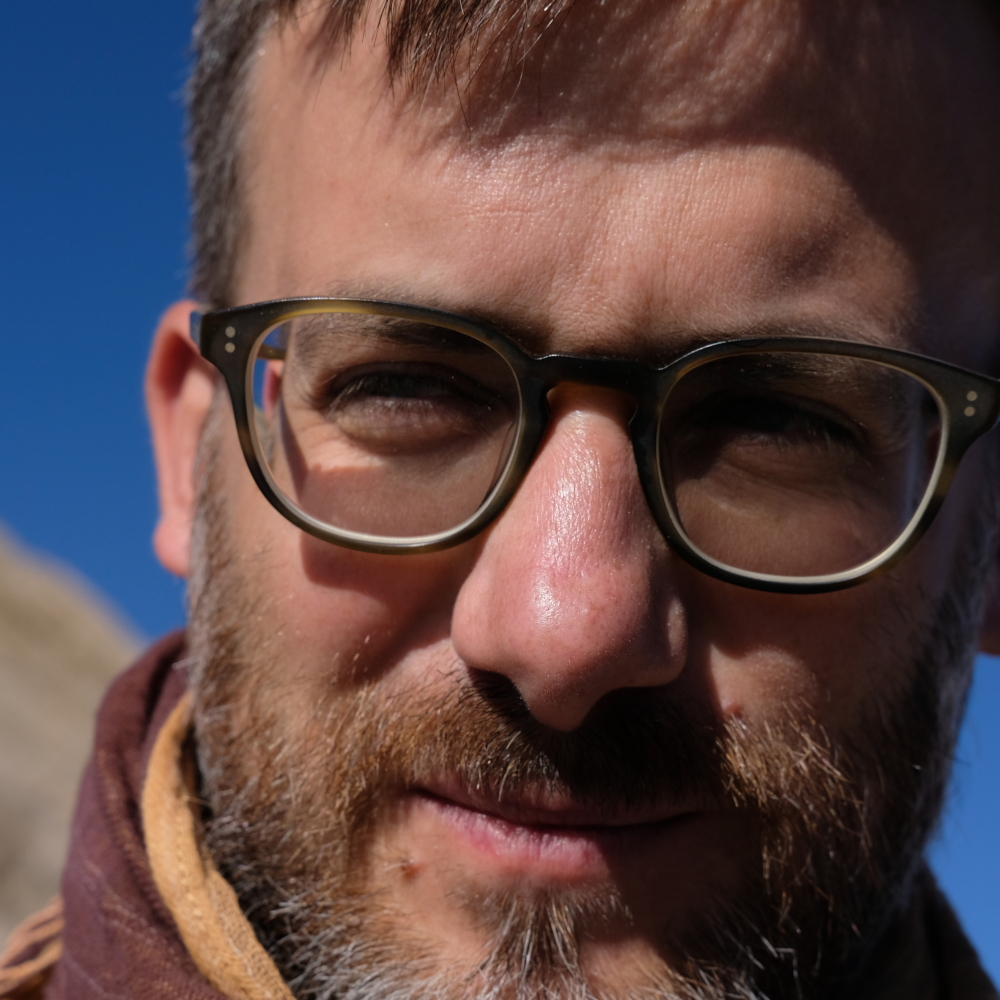}}]{Matteo Brancaleoni} is a Senior Staff Engineer at VoiSmart Srl, Italy, where he develops unified communication systems with a focus on VoIP infrastructures and videoconferencing platforms. His expertise includes real-time communications, telephony protocols, WebRTC, and distributed architectures. His research and professional activities center on designing scalable, reliable communication frameworks and exploring their extension with artificial intelligence. He is particularly interested in protocol design, communication middleware, and AI integration into real-time distributed systems. He can be contacted at \url{matteo.brancaleoni@voismart.it}
\end{IEEEbiography}
\vspace*{-1.45cm}
\begin{IEEEbiography}[{\includegraphics[width=1in,height=1.25in,clip,keepaspectratio]{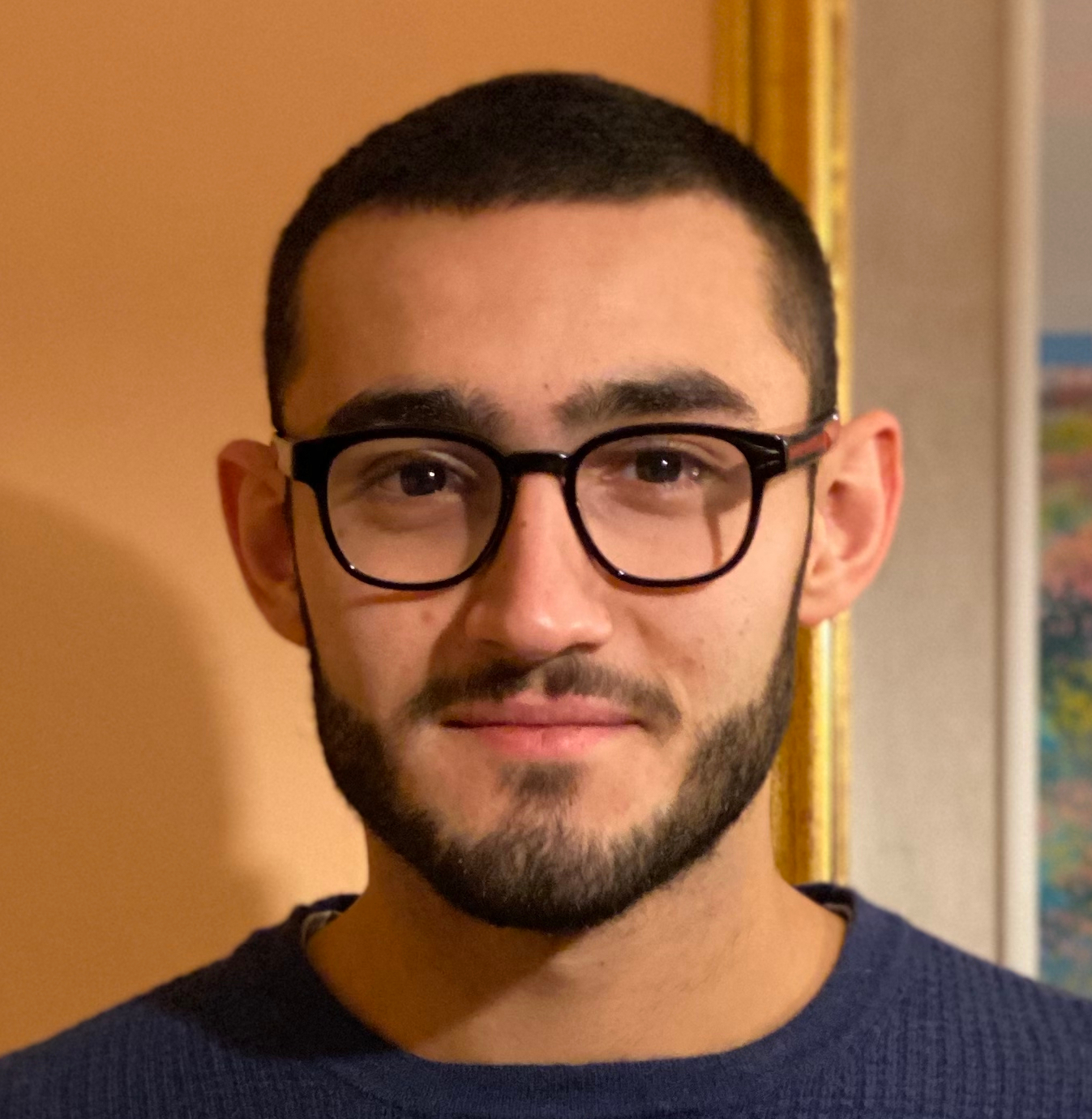}}]{Dario Pellegrino} is currently a Master of Science student in Computer Science. He obtained his Bachelor's degree in Computer Science with a thesis project developed in collaboration with the ADAPT Lab and VoiSmart, focusing on the SWIM project. Recently, he started working as a Junior Software Engineer at VoiSmart, contributing to research and development activities on AI-based digital agents and assistants in the domain of telephony and PBX systems. His main research interests include software engineering and artificial intelligence, with a particular enthusiasm for both low-level aspects such as machine learning and the application of AI to software engineering. He can be contacted at \url{dario.pellegrino@voismart.it}
\end{IEEEbiography}
\newpage
\end{document}

%% file: sects/abstract.tex
\begin{abstract}
   Real-time automatic speech recognition systems are increasingly integrated into interactive applications, from voice assistants to live transcription services. However, scaling these systems to support multiple concurrent clients while maintaining low latency and high accuracy remains a major challenge.
   In this work, we present \ourtool{SWIM}, a novel real-time ASR system built on top of OpenAI's \texttt{Whisper} model that enables true model-level parallelization for scalable, multilingual transcription.\ \ourtool{SWIM} supports multiple concurrent audio streams without modifying the underlying model. It introduces a buffer merging strategy that maintains transcription fidelity while ensuring efficient resource usage.
   We evaluate \ourtool{SWIM} in multi-client settings---scaling up to 20 concurrent users---and show that it delivers accurate real-time transcriptions in English, Italian, and Spanish, while maintaining low latency and high throughput. While \texttt{Whisper-Streaming} achieves a word error rate of approximately 8.2\% with an average delay of approximately 3.4\,s in a single-client, English-only setting, \ourtool{SWIM} extends this capability to multilingual, multi-client environments. It maintains comparable accuracy with significantly lower delay---around 2.4\,s with 5 clients---and continues to scale effectively up to 20 concurrent clients without degrading transcription quality and increasing overall throughput. Our approach advances scalable ASR by improving robustness and efficiency in dynamic, multi-user environments.
\end{abstract}

%% file: sects/introduction.tex
\section{Introduction}\label{sec:intro}
\noindent\IEEEPARstart{A}{utomatic} speech recognition (ASR) systems\footnote{From here on, we use the term \textit{system} to refer specifically to a software system, typically in a client-server setup, where the client is the user and the server hosts the ASR model.} are now ubiquitous in everyday applications~\cite{Sharma16, Ping18, Alharbi21, Lecouteux11}, ranging from voice assistants for human-computer interaction~\cite{Ram18, Gerosa09} to transcription services~\cite{Tjandra17, Dubey22}.
In recent years, ASR has advanced significantly through deep learning~\cite{Hannun14, Prabhavalkar23, Li22, VanDenOord16}.
The field has shifted from traditional \textit{cascaded} architectures~\cite{Furui05, Gales08}, which separate the recognition pipeline into distinct stages, to \textit{end-to-end} models~\cite{Aldarmaki22}, where a single neural network jointly learns the entire task~\cite{Baevski20, Chen21c}.
\texttt{Whisper}, developed by OpenAI, is a notable transformer-based end-to-end ASR model~\cite{Radford23}. It represents a well-balanced end-to-end solution in the ASR landscape, offering a trade-off between computational cost, speed, and transcription accuracy, as demonstrated by~\mbox{\citet{Radford23}.}

\vspace{0.5em}\noindent\textbf{ASR Systems and Their Challenges.}\quad
Contemporary ASR systems, particularly those based on conversational AI, are generally characterized along two dimensions: \textit{single-client} systems~\cite{Hannun14, Amodei16}, which serve a single user at a time, and \textit{multi-client} systems, which concurrently serve multiple users~\cite{Morris01, Li19}. These architectures are further classified based on deployment, into \textit{local} systems, which operate on the user's device, and \textit{remote} systems, which are hosted on servers and accessed via network connections.
Recent trends in ASR research have increasingly focused on real-time applications, where both low latency and high throughput are critical~\cite{Sharma16, Garner09, Arriaga24, Ivanov16}. In these settings, the system must generate responses within a few seconds to meet users' expectations for immediate feedback. Although local ASR systems offer enhanced privacy~\cite{Ahmed20a, Abdullah21}, their performance is inherently constrained by the limited computational power and memory of user devices~\cite{Georgescu21}, as well as their susceptibility to adverse environmental conditions. For example, in audio-only communication channels such as \textit{voice over internet protocol} (VoIP) and telephone calls, audio signals are typically routed to centralized remote ASR systems that benefit from greater computational resources and cloud-based services~\cite{Ramirez24, Zhu22}.

\vspace{0.5em}\noindent\textbf{Single-client vs. Multi-client Systems.}\quad
Furthermore, single-client systems are not without limitations~\cite{Basak22}. E.g., \texttt{Whispy}~\cite{Bevilacqua24} and \texttt{Whisper-Streaming}~\cite{Machacek23} lack model-level parallelization, limiting their scalability in multi-audio streams scenarios. In contrast, remote ASR systems can leverage greater computational resources and memory, enabling efficient scaling to support multiple clients concurrently~\cite{Polak22}. Such systems benefit from optimized resource allocation, reduced latency via parallel processing, and consistent performance across diverse user populations~\cite{Pratap20, Cui21}.

Nevertheless, scaling real-time ASR to handle many concurrent clients remains a major challenge. This is largely due to the computational demands of real-time models, which must deliver low-latency transcriptions without losing accuracy---a critical requirement in applications like voice assistants~\cite{Zhang19c}, live captioning~\cite{Kuhn24}, and safety-critical domains such as healthcare~\cite{Jamal17, Mustafa15, Latif21} or legal proceedings~\cite{Loakes22, Komter22}.

\vspace{0.5em}\noindent\textbf{Contributions.}\quad
In this paper, we introduce \ourtool{SWIM} (\textbf{S}erve \textbf{W}hisper \textbf{I}n \textbf{M}ulti-client), a novel real-time ASR system built  on top of the \texttt{Whisper} model, designed to serve multiple clients simultaneously while maintaining low latency and high throughput.\ \ourtool{SWIM} prioritizes scalability, enabling efficient handling of multilingual audio streams under varying workloads. Unlike \texttt{Whisper-Streaming}~\cite{Machacek23} and \texttt{Whispy}~\cite{Bevilacqua24}, \ourtool{SWIM} achieves true model-level parallelization, ensuring stable performance even under high-load conditions. Through empirical evaluation, we demonstrate that a single \texttt{Whisper} instance running \ourtool{SWIM} can transcribe multiple multi-lingual audio streams in parallel, maintaining optimal accuracy in real-time scenarios.
Our evaluation spans various multi-client settings (e.g., 5, 10, and 20 concurrent users) and includes: \begin{inparaenum}[(i)]
   \item assessment of real-time playback performance, in terms of delay with respect to wall-clock time,
   \item measurement of transcription quality using \textit{word error rate} and accuracy metrics, and
   \item evaluation of the system's handling of multilingual inputs, including English, Italian, and Spanish.
\end{inparaenum}
This work tackles scalability challenges in real-time ASR, advancing the state of the art in multi-client speech recognition for dynamic environments.

Our main contributions are:
\begin{enumerate}
   \item a scalable real-time ASR system enabling model-level parallelization for concurrent multilingual transcription;
   \item a buffer merging strategy that maintains transcription accuracy across clients without added latency; and
   \item a demonstration of \texttt{Whisper}'s unmodified ability to handle multiple parallel audio streams from diverse sources effectively.
\end{enumerate}

The remainder of the paper is organized as follows.
Sect.~\ref{sec:bg} provides background.
Sect.~\ref{sec:state-of-the-art} reviews real-time ASR systems, focusing on \texttt{Whisper} and its limitations;
Sect.~\ref{sec:design} details the design of \ourtool{SWIM}, and Sect.~\ref{sec:evaluation} presents our evaluation, and Sect.~\ref{sec:conclusion} concludes the paper.

%% file: sects/background.tex
\section{Background}\label{sec:bg}
We provide background on speech-to-text ASR models, the \texttt{Whisper} Model, and the \texttt{Whisper-Streaming} system.

\vspace{0.5em}\noindent\textbf{Speech-to-Text ASR Models.}\quad
ASR converts spoken language into written text using machine learning.
Traditionally, ASR systems have relied on hidden Markov models~\cite{Rabiner89, Gales08} and Gaussian mixture models~\cite{Furui05}, which dominated the field for many years.
ASR models vary from high-performance, \textit{heavy} systems designed for offline processing~\cite{Baevski20, Gulati20} to \textit{lightweight} versions optimized for real-time use on limited hardware~\cite{He19, Hoy18}. These differences shape key architectural and optimization decisions.

Traditional ASR models often used a \textit{cascaded} architecture~\cite{Furui05, Gales08}, where separately trained components process speech sequentially. In contrast, \textit{end-to-end} models~\cite{Aldarmaki22} typically integrate these stages into a single neural network trained jointly~\cite{Baevski20, Chen21c}, simplifying the pipeline and often improving performance, especially with large datasets~\cite{Baevski20, Chen21c}.
A prominent example is the \textit{transformer} architecture~\cite{Baevski20, Han21}, which replaces recurrent and convolutional neural networks with self-attention mechanisms. Transformers enable high parallelization and achieve state-of-the-art results across many ASR tasks~\cite{Baevski20, Vaswani17, Pratap24}.
ASR models are commonly trained on large-scale datasets such as LibriSpeech~\cite{Panayotov15}---available in variants like \begin{inparaenum}
   \item LibriSpeech Long~\cite{Park25} and
   \item Multilingual LibriSpeech~\cite{Pratap20b}
\end{inparaenum}---as well as Fleurs~\cite{Conneau22}. These datasets provide thousands of hours of transcribed speech, supporting robust model training.

Performance is typically measured by \textit{word error rate} (WER)~\cite{Ali18, Klakow02, Wang03}, \textit{character error rate} (CER)~\cite{Guyon98, Park24b, Sawata22}, and \textit{sentence error rate} (SER)~\cite{Spiccia16, Raybaud11}, which quantify transcription accuracy using normalized Levenshtein distance~\cite{Li07, Heeringa04} between predicted and reference texts.
Synchronized transcription of audio streams is crucial---especially for subtitles, captions, or live transcription~\cite{Shi21, Han22}. To address this, ASR systems segment audio into temporal segments transcribed individually. This \textit{timestamp-aligning}\footnote{Also known as \textit{temporal-}, \textit{forced-}, or \textit{word-level alignment}.} process~\cite{Li22b, Chu15} aligns transcriptions with the audio, enabling extraction of precise temporal segments.

\vspace{0.5em}\noindent\textbf{Whisper Model.}\quad
\texttt{Whisper}~\cite{Radford23} is an open-source ASR model developed by OpenAI\@. It is trained on a large corpus of multilingual and multitask datasets, enabling it to perform multilingual speech recognition.\ \texttt{Whisper} is a prime example of an end-to-end ASR model based on the Transformer architecture.
Unlike traditional ASR models, \texttt{Whisper} processes raw audio and uses a sequence-to-sequence transformer model to generate text.
Being an end-to-end model, the \texttt{Whisper} consists of an \textit{encoder} and a \textit{decoder} architecture. The former processes the audio input into a dense representation, while the latter translates the representation into a sequence of text tokens using cross-attention mechanisms.
Beyond speech-to-text transcription, \texttt{Whisper} demonstrates high proficiency in additional tasks, including:
\begin{inparaenum}
   \item speech translation,
   \item language identification and
   \item voice activity detection.
\end{inparaenum}
At inference time, a list of start and end timestamps can be provided to specify audio segments, bypassing voice activity detection by assuming these intervals contain speech.
\texttt{Whisper} is available in several sizes and variants---e.g., \texttt{tiny}, \texttt{small}, \texttt{medium}, \texttt{large-v3}. Recently, a \texttt{turbo} version of the pre-trained \texttt{large-v3}, pruned and fine-tuned, was introduced, offering a 5--8\(\times\) speedup with minimal accuracy loss.

\section{State-of-the-Art}\label{sec:state-of-the-art}
\noindent\textbf{Real-time Whisper.}\quad
\texttt{Faster-whisper}\footnote{\label{foot:faster-whisper}\url{https://github.com/guillaumekln/faster-whisper}} is a reimplementation of \texttt{Whisper} built on top of \texttt{CTranslate2}.\footnote{\url{https://github.com/OpenNMT/CTranslate2}} It has been widely adopted by several real-time ASR systems, including \texttt{Whispy}~\cite{Bevilacqua24} and \texttt{Whisper-Streaming}~\cite{Machacek23}.

\texttt{Whisper-Streaming} enables real-time speech transcription and translation by using an incremental buffer and local agreement mechanism to minimize latency and maximize throughput~\cite{Machacek23}.
\texttt{Whispy}, on the other hand, is designed to support real-time use of \texttt{Whisper} by employing Levenshtein distance to align overlapping transcriptions and select the most accurate segments~\cite{Bevilacqua24}.
Despite adopting different approaches, both projects rely on \texttt{faster-whisper} as the \texttt{Whisper} implementation, minimizing latency and offering a valuable contribution to the field. However, \texttt{Whisper}---including its variants---is GPU bound and introduce big bottlenecks when it comes to real-time scenarios and parallelization.
Other projects have proposed solutions for these issues, such as \texttt{Whisper-T}~\cite{Wang24c}.\ \texttt{Whisper-T} aims to reduce \texttt{Whisper}'s GPU load and power consumption through lightweight decoding strategies, beam pruning, and CPU/GPU pipelining. Nevertheless, the project remains limited to a local environment, processing only one audio stream at a time. Achieving significant scalability and parallelization for \texttt{Whisper} remains an open challenge and is essential for building remote, multi-client ASR systems.

\vspace{0.5em}\noindent\textbf{Limitations of the State-of-the-art.}\quad
The goal is to develop a system capable of handling multiple requests while running on relatively modest hardware (e.g., a few GPUs or even a single GPU).
Achieving this presents several challenges, including:
\begin{inparaenum}
   \item most existing \texttt{Whisper} implementations (except \texttt{faster-whisper}) exhibit significant performance degradation in real-time scenarios, and
   \item the absence of model-level parallelization support.
\end{inparaenum}
Several \texttt{Whisper} implementations, such as: \texttt{insanely-fast-whisper}\footnote{\url{https://github.com/Vaibhavs10/insanely-fast-whisper}} and \texttt{WhisperX}~\cite{Bain23} (which builds upon \texttt{faster-whisper}), are not suitable for real-time applications due to latency issues.
The former performs poorly on short audio clips (e.g., 15–30 seconds) and is not designed for processing raw audio buffers. While it includes a \textit{diarization}\footnote{%
\textit{Diarization} refers to segmenting an audio stream by speaker identity, allowing the system to identify who spoke when.
} feature, the latter inherits the limitations of \texttt{faster-whisper}, making it slower than its base implementation.
Moreover, \texttt{Whisper} and its variants are not inherently designed to process multiple audio streams concurrently, primarily due to two reasons:
\begin{inparaenum}
   \item although available VRAM may support multiple model instances, GPU compute resources are quickly exhausted during inference, and
   \item when multiple \texttt{Whisper} instances are launched, only one can utilize GPU computation at a time, forcing the others to wait---resulting in serialized audio processing.
\end{inparaenum}
A high-cost solution, feasible for large companies, involves horizontal scaling with a dedicated GPU for each \texttt{Whisper} instance.
A more affordable alternative is asynchronous processing on a single GPU, though this is poorly supported in \texttt{faster-whisper} and similar implementations.

%% file: sects/design.tex
\begin{figure*}[t]
   \centering
   \includegraphics[width=\linewidth]{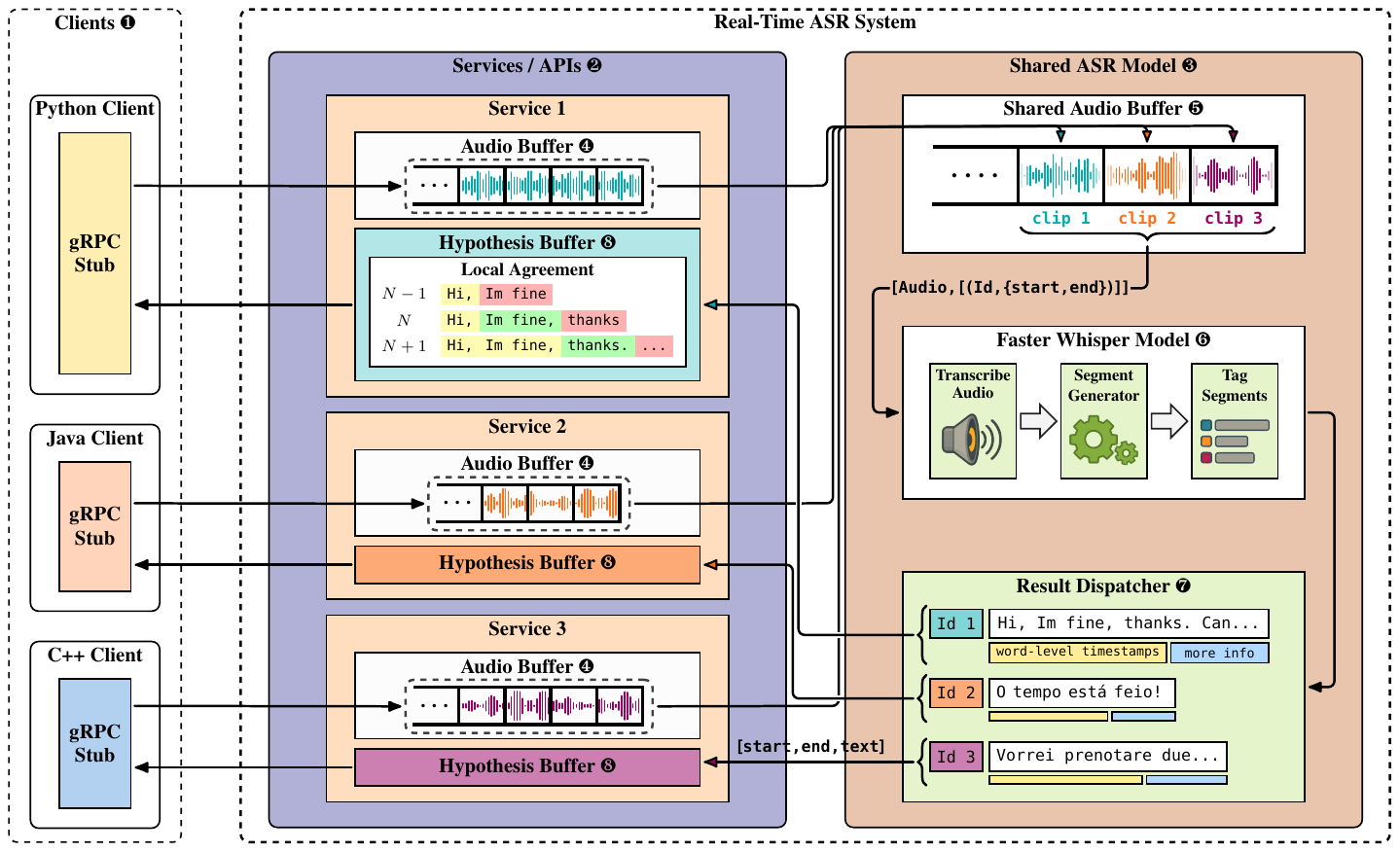}
   \caption{Architecture of \ourtool{SWIM}. Multiple clients \ding{182} stream audio to dedicated \textit{services} \ding{183}, which forward data to a \textit{shared ASR model} \ding{184} for parallel processing.}%
   \label{fig:architecture}
\end{figure*}

\section{A Deep Dive into SWIM}\label{sec:design}
\ourtool{SWIM} is a multi-client, real-time ASR system built on top of the \texttt{Whisper} model. It is designed to: \begin{inparaenum}
   \item concurrently process multiple multilingual real-time audio streams,
   \item maintain low latency and high throughput, and
   \item deliver accurate transcriptions for all clients.
\end{inparaenum}
\subsection{Overview}\label{sec:overview}
The actors in \ourtool{SWIM}'s architecture are illustrated in Fig.~\ref{fig:architecture}.
The system consists of multiple clients \ding{182}, each streaming audio to a dedicated service \ding{183}. Each service is responsible for two main tasks:
\begin{inparaenum}
   \item building an \textit{audio buffer} from incoming audio chunks and submitting it to the \textit{shared ASR model} \ding{184} for transcription, and
   \item maintaining a \textit{hypothesis buffer} \ding{189}, which stores intermediate transcriptions and applies a \textit{local agreement} strategy to ensure reliability.
\end{inparaenum}
The \textit{shared ASR model} assembles clips from all \textit{audio buffers} \ding{186}, passes them to a \texttt{faster-whisper} instance \ding{187} for inference, and relays the transcriptions to the \textit{result dispatcher} \ding{188}.

\subsection{Clients}\label{sec:clients}
\ourtool{SWIM} is designed to be a multi-client ASR system. It can be deployed on a remote server or adapted for local use without the client-server setup---e.g., handling multiple audio sources on a single machine.
Clients, shown in Fig.~\ref{fig:architecture}\raisebox{0.1em}{-}\ding{182}, send audio streams to the system. These can be any audio-capturing devices, such as smartphones, computers, or single-board development kits like \texttt{Raspberry Pi} or \texttt{Arduino}.
Client implementations emphasize polyglotism and platform agnosticism, supporting any programming language compatible with the \texttt{gRPC} protocol.\footnote{\url{https://grpc.io}}
Clients receive transcriptions of their audio streams via a \textit{bidirectional} communication channel, allowing the server to return transcriptions and other related information.
Clients must send audio streams sampled at 16\,kHz, the rate required by \texttt{Whisper} and widely adopted in modern ASR systems. Knowing this sampling rate is essential for the system to effectively process audio buffers---such as chunking based on timestamps or synchronizing local transcriptions.

\subsection{Services}\label{sec:services}
As illustrated in Fig.~\ref{fig:architecture}\raisebox{0.1em}{-}\ding{183}, services handle the audio streams sent by clients. Each service is dedicated to a single \texttt{gRPC} client and constructs an \textit{audio buffer} from the incoming chunks.
These \textit{audio buffers}---shown in \tikz\fill[MyTealBlue] (0,0) circle (0.1cm);, \tikz\fill[MyOrange] (0,0) circle (0.1cm);, and \tikz\fill[MyBloodRed] (0,0) circle (0.1cm); in Fig.~\ref{fig:architecture}\raisebox{0.1em}{-}\ding{183}---are continuously forwarded to the \texttt{Whisper} model for processing (indicated by arrows from \ding{183} to \ding{184}) to maintain real-time performance.
To reduce processing overhead, the audio buffer is kept as short as possible (typically 10--15 seconds). If the buffer exceeds its maximum duration (e.g., 15 seconds), the oldest content is trimmed. This trimming applies only to the already transcribed portion of the buffer, which has been marked as \textit{confirmed} by the \textit{local agreement} mechanism within the \textit{hypothesis buffer}\footnote{%
   The \textit{hypothesis buffer} (Fig.~\ref{fig:architecture}\raisebox{0.1em}{-}\ding{189}) stores transcriptions from the shared model's result dispatcher (Fig.~\ref{fig:architecture}\raisebox{0.1em}{-}\ding{188}). It functions by comparing the latest confirmed transcription with the current one.
}
(see Sect.~\ref{sec:hypothesis-buffer}).
Thanks to the \textit{word-level} timestamp alignment provided by \texttt{faster-whisper},
the buffer can be segmented at precise word boundaries, preventing word splits. Each segment—depicted in Fig.~\ref{fig:architecture}\raisebox{0.1em}{-}\ding{188}—is a continuous sequence of words with start/end timestamps, along with metadata such as full text and detected language.
This design facilitates operations like overlap resolution, insertion, and deletion in the \textit{hypothesis buffer} (\ding{189}\raisebox{0.1em}{-}Fig.~\ref{fig:architecture}).
Specifically, the buffer tail (\ding{185}\raisebox{0.1em}{-}Fig.~\ref{fig:architecture}) is truncated at the timestamp of the first confirmed word that occurs before the midpoint of the buffer duration, i.e., $\ell < \frac{B}{2}$, where $B$ is the total buffer length (e.g., 15 seconds).
As shown in Fig.~\ref{fig:architecture}\raisebox{0.1em}{-}\ding{183} (Service 1), the confirmed portion of the \textit{hypothesis buffer} is marked in \textit{yellow} \tikz\fill[yellow] (0,0) circle (0.1cm); and \textit{green} \tikz\fill[green] (0,0) circle (0.1cm); while the unconfirmed portion is shown in \textit{red} \tikz\fill[red] (0,0) circle (0.1cm);. When a new text segment is confirmed (as in Service 1), it is marked \textit{green} and sent back to the client, along with its start and end timestamps.
Details on the local agreement policy can be found in Sect.~\ref{sec:hypothesis-buffer}.

\subsection{Shared ASR Model}\label{sec:shared-asr-model}

\vspace{0.5em}\noindent\textbf{Shared Audio Buffer.}\quad
As we have already mentioned, \ourtool{SWIM} is designed to be a multi-client ASR system. However, each service does not have a dedicated instance of the \texttt{Whisper} model. Instead, \ourtool{SWIM} is based on a \textit{shared ASR model}\footnote{
   When we refer to the \textit{Shared ASR model} or \textit{Shared model}, we mean an instance of a custom object that wraps the \texttt{faster-whisper} model and provides a set of methods to interact with it by allowing it to be used in a multi-client environment.
} Fig.~\ref{fig:architecture}\raisebox{0.1em}{-}\ding{184},
which, as the name suggests, is shared across all services\hspace{0.3em}\ding{183}. To function properly and to enable parallel processing of multiple audio streams, the \textit{shared model} is designed to operate using a single \textit{shared audio buffer} Fig.~\ref{fig:architecture}\raisebox{0.1em}{-}\ding{186}, which is likewise shared among all services. The \textit{shared ASR model} is responsible of orchestrating the parallel progressing of the audio streams, by continuously: \begin{inparaenum}
   \item gathering the services \textit{audio buffers} into a \textit{shared audio buffer},
   \item processing them in parallel using a \texttt{faster-whisper} instance Fig.~\ref{fig:architecture}\raisebox{0.1em}{-}\ding{187}, and
   \item tagging and sending the text results back to the services (see Result Dispatcher Fig.~\ref{fig:architecture}\raisebox{0.1em}{-}\ding{188}).
\end{inparaenum}

The \textit{shared audio buffer} is a monolithic raw buffer built on top of the individual \textit{audio buffers} managed by each dedicated service.
Its internal structure enables the system to track which portions of the buffer belong to each client. Alongside the \textit{audio buffer}, the system maintains a list of pairs consisting of: \begin{inparaenum}
   \item the unique identifier of the associated service, and
   \item the clip timestamps marking the start and end of the corresponding service's audio segment.
\end{inparaenum}
The \textit{shared audio buffer} is populated by the \textit{shared ASR model} when all the services are ready to process their \textit{audio buffer}---that is, a progressing request (e.g., a novel audio chunk) has been received from the client.
Several policy mechanisms to avoid deadlocks could be easily implemented to handle the situation when one or more services are not ready to process their \textit{audio buffer}.
For instance, the system could wait for a certain amount of time (e.g., range from 1 to 2 seconds) for a service to be ready. When unready, the system could skip it and proceed regardless with two behaviors: \begin{inparaenum}
   \item skipping the service, but pausing or kicking it out from the system, and
   \item skipping the service only for the current transcription, but keeping it active for the next ones.
\end{inparaenum}

Each service operates autonomously and processes its own audio stream independently, without interference from other services. As a result, synchronization issues are implicitly managed by the system architecture.
A core assumption of \ourtool{SWIM} is that the audio stream is persistent and transmitted continuously in fixed-duration chunks, with each chunk having a duration \(D \in (0, 1]\). Consequently, as Fig.~\ref{fig:maxdelay} illustrates, there always exist two consecutive audio chunks, denoted by \((D_1, T_1)\) and \((D_2, T_2)\), where \( D_j \) represents the audio chunk received by client \(C_j\) at time \(T_j\). 
Let \((T, i) \leftarrow (\max\{T_1, T_2\}, \arg\max_{i \in \{0, 1\}} \{T_i\})\), then the following inequality holds:
\[
   T - T_{|i - 1|} - \varepsilon \leq 1s
\]
where:
\begin{compactenum}
   \item \( T_{|i - 1|} \) is the timestamp of the last audio chunk received by client \( C_{|i - 1|} \), and
   \item \( \varepsilon = o(1) \) represent the network latency of \(T\).
\end{compactenum}

This architectural assumption, as Fig.~\ref{fig:maxdelay} depicts, ensures that the maximum delay incurred while waiting for all services to be ready is bounded by
\[
   \max(0, 1 - T_p) < 1s,
\]
where \( T_p \) is the processing time of the transcriptions.
Intuitively, this represents the worst-case scenario where one service has to wait for the other to finish processing and sending its audio chunk.
\begin{center}
   \includegraphics[width=\linewidth]{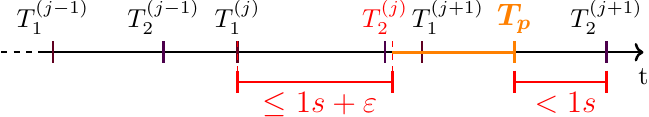}
   \captionof{figure}{Audio chunk processing time diagram.}%
   \label{fig:maxdelay}
\end{center}
The outer maximum with zero ensures that we do not consider negative delays: if the \textit{shared ASR} takes longer than 1 second to process all active services, the next chunk will already have arrived for every connected services, so there's no extra waiting time.

\vspace{0.5em}\noindent\textbf{Faster \texttt{Whisper} Model.}\quad
In the adopted architecture, the entire \textit{shared audio buffer} is passed to the model as a single, continuous waveform. Individual client contributions are identified through a list of uniquely associated identifiers and temporal intervals defined by a pair \texttt{(start, end)}---denoted as \texttt{Timestamp} in Fig.~\ref{fig:architecture} (in the path from \ding{186} to \ding{187}).

Inference is then performed on the entire \textit{shared} buffer, based on the artificially defined clips---that is, the \textit{shared audio buffer} divides the audio stream into clips, each associated with a unique client identifier.
For each registered clip---denoted by \tikz\fill[MyTealBlue] (0,0) circle (0.1cm);, \tikz\fill[MyOrange] (0,0) circle (0.1cm);, \tikz\fill[MyBloodRed] (0,0) circle (0.1cm); colors in Fig.~\ref{fig:architecture}\raisebox{0.1em}{-}\ding{186}--- the system extracts a corresponding transcription segment. As briefly mentioned in Sect.~\ref{sec:services}, the length of the service \textit{audio buffer} should range between 10 and 15 seconds. This design choice is motivated by multiple, complementary factors. First, it alleviates the computational load by limiting the input size processed at each inference step.
Second, and more importantly, clips that are too long—exceeding approximately \(30\) seconds are cut off the excessive part of the clip.

\vspace{0.5em}\noindent\textbf{Result Dispatcher.}\quad
After inference, the text segments are returned in the exact order in which they were processed. With this design choice, the system can easily associate each segment with the corresponding client identifier.
Before returning the results, the system adjusts the segment timestamps---including word-level timings---so they align with the relative time of the original service audio buffer. This adjustment is necessary because the timestamps are relative to the \textit{shared audio buffer}, not to the service-specific \textit{audio buffer}.
Finally, the segment results are forwarded to the corresponding service, which processes the segment and delivers the transcription back to the client according to the previously described mechanism (on the arrow from \ding{188} to \ding{189}).

\subsection{Hypothesis Buffer with Local Agreement}\label{sec:hypothesis-buffer}

\ourtool{SWIM} uses an adapted \textit{hypothesis buffer}, inspired by the mechanism introduced in \texttt{Whisper-Streaming}~\cite{Machacek23}, which implements a \textit{local agreement} strategy to ensure the reliability of the transcriptions produced by the \texttt{Whisper} model. Our implementation employs a \textit{similarity metric}~\cite{Bevilacqua24, Li07, Navarro01} rather than strict token equality to govern the behavior of the \textit{hypothesis buffer}.
\ourtool{SWIM} uses the \texttt{QRatio},\footnote{
   Later in the paper we refer to this \texttt{QRatio} metric also as \textit{similarity metric}, \textit{similarity ratio}, or \textit{similarity score}.
} also known as normalized Indel distance, introduced by Bachmann, defined as:
\begin{equation*}
   QRatio(s1, s2) = \left(1 - \frac{d(s1, s2)}{|s1| + |s2|}\right) \times 100,
\end{equation*}
where \(d(s1, s2)\) is the Indel distance~\cite{Deza09, Sellers74} between the two strings \(s1\) and \(s2\)---a variant of the Levenshtein distance~\cite{Levenshtein66} where substitutions have a cost of two---, and \(|s1|\) and \(|s2|\) are the lengths of the two strings.

The motivation behind these changes is twofold: \begin{inparaenum}
   \item to account for punctuation, apostrophes, or minor discrepancies---whether introduced by \texttt{Whisper} itself or by individual services---and
   \item to populate the local agreement list (a buffer of words along with timestamps \texttt{(start, end, text)}) and retain content that may have been slightly altered by the model.
\end{inparaenum}

When a new segment is received, the system replaces the previous segment with the new one, inserting the first $n$ consecutive tokens whose similarity ratio exceeds or equals 98\% compared to the corresponding $n$ tokens from the previous segment. The purpose of this insertion mechanism is to find overlaps between the two segments, which helps hypothesis buffer with being more accurate and reliable.
The local agreement mechanism enables token confirmation by comparing the new segment against the previous one, using a token-by-token equality check. By setting a confirmation threshold at a 95\% similarity ratio, the mechanism accepts variations such as added or removed punctuation without wasting additional iterations, while maintaining high transcription accuracy. An illustrative example is shown in the local agreement box under the first \ding{189} of Fig.~\ref{fig:architecture}.

If no confirmation occurs via the local agreement policy, a fallback strategy is triggered before proceeding to the next iteration without any confirmation.
The system takes the first half of the previous transcription---based on word count---and generates all possible ordered prefixes. These prefixes are then compared to those of the new transcription, but only among tokens whose end timestamps fall within the duration of the selected half of the previous segment. This constraint ensures that matching is confined to the relevant portion of the new transcription, preventing premature confirmation of content beyond the midpoint of the previous segment.
The prefix with the highest similarity score is selected, and all its constituent tokens are confirmed individually, as confirmation is performed on a token-by-token basis, adopting the same similarity metric as in the local agreement.
This fallback mechanism helps recover from shifts in phrasing or tokenization that can put the service in a never confirming state until the service audio buffer gets trimmed.

%% file: sects/evaluation.tex
\section{Evaluation}\label{sec:evaluation}
The evaluation of \ourtool{SWIM} utilizes four datasets: the \textit{Google Fleurs dataset}~\cite{Conneau22}, the \textit{Multilingual LibriSpeech dataset}~\cite{Pratap20b}, the \textit{LibriSpeech-Long dataset}~\cite{Park25}, and the \textit{Italian Parkinson's Voice dataset}~\cite{Dimauro16, Dimauro17, Dimauro19}. We use \texttt{large-v3-turbo}, a pruned and fine-tuned version of the pre-trained \texttt{large-v3} model, which offers significant improvements in efficiency. A replication package for the experiments is available at:
\begin{center}\url{https://doi.org/10.5281/zenodo.15856828}\end{center}

\subsection{Datasets}
\textbf{Description.}\quad
The \textit{Google Fleurs dataset} offers a diverse multilingual benchmark with recordings from 102 languages, making it ideal for evaluating \ourtool{SWIM}'s robustness across a wide range of speech types.
The \textit{Multilingual LibriSpeech dataset} consists of read audiobooks in 8 European languages, supporting evaluation in moderately resourced multilingual settings.
The \textit{LibriSpeech-Long dataset} extends LibriSpeech by including longer audio segments, testing the model's ability for long-range temporal modeling.
Finally, the \textit{Italian Parkinson's Voice dataset} contains Italian speech samples from Parkinson's disease patients and control subjects, enabling assessment of \ourtool{SWIM} in pathological speech recognition scenarios, which is critical for evaluating performance on impaired speech.\smallskip

\noindent\textbf{Preparation.}\quad
We preprocess all datasets to align audio files precisely with their transcriptions. Using the \texttt{Whisper large-v3} model~\cite{Radford23}, we obtain word-level ASR alignments that map audio to text accurately. This precise word-level timestamping is essential for evaluating \ourtool{SWIM} in real-time ASR scenarios requiring tight audio-text synchronization.

We chose the \texttt{Whisper large-v3} model for its balance of latency, accuracy, and speed, making it well suited for alignment. Its state-of-the-art performance also provides a strong baseline for evaluating \ourtool{SWIM}, given our method's constraints and the \texttt{large-v3 turbo} variant's limitations.

The prepared datasets support evaluation of \ourtool{SWIM} across multiple scenarios:
\begin{inparaenum}
	\item the \textit{Multilingual LibriSpeech} and \textit{Google Fleurs} datasets assess multilingual concurrent performance, with average audio lengths of approximately 15 and 10 seconds, respectively;
	\item the \textit{LibriSpeech-Long} dataset evaluates long-form audio processing, with an average duration around 5 minutes; and
	\item the \textit{Italian Parkinson's Voice} dataset offers an intermediate scenario, with audio segments around 1 minute long, featuring recordings used for disease diagnosis and including specific Italian regional accents.
\end{inparaenum}

\subsection{Experimental Setup}
We conduct our experiments on a server equipped with: \begin{inparaenum}
	\item an NVIDIA RTX 6000 Ada Generation GPU with 48GB of VRAM,
	\item Intel\,\textcopyright{} Xeon\,\textcopyright{} Gold 5412U CPU with 24 cores and 48 threads, 
	\item 128GB of RAM, and
	\item the operating system is Ubuntu 22.04 LTS\@.
\end{inparaenum}

\begin{figure*}[htbp]
	\centering

	\begin{adjustbox}{width=0.95\textwidth}
		\begin{minipage}{\textwidth}

			\ \vspace*{0.2cm}
			\begin{subfigure}[b]{\textwidth}
				\centering
				\captionsetup{skip=1pt}
				\includegraphics[width=\textwidth]{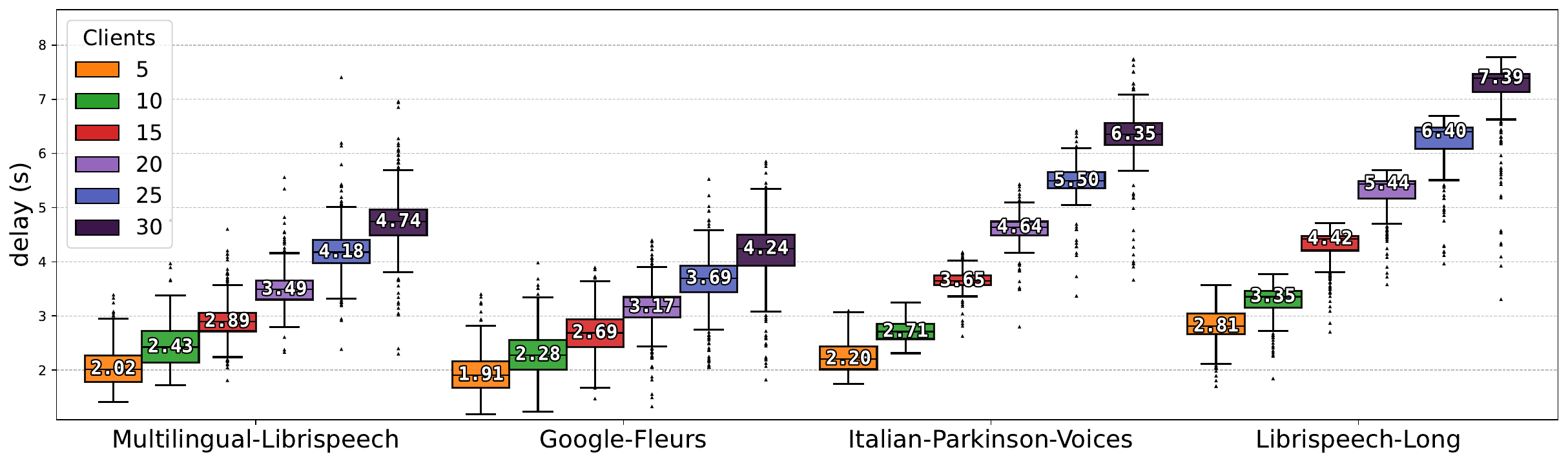}
				\caption{Chunk duration of 1.0s}%
				\label{fig:evaluation-delay1.0}
			\end{subfigure}
			\ \vspace*{0.2cm}
			\begin{subfigure}[b]{\textwidth}
				\centering
				\captionsetup{skip=1pt}
				\includegraphics[width=\textwidth]{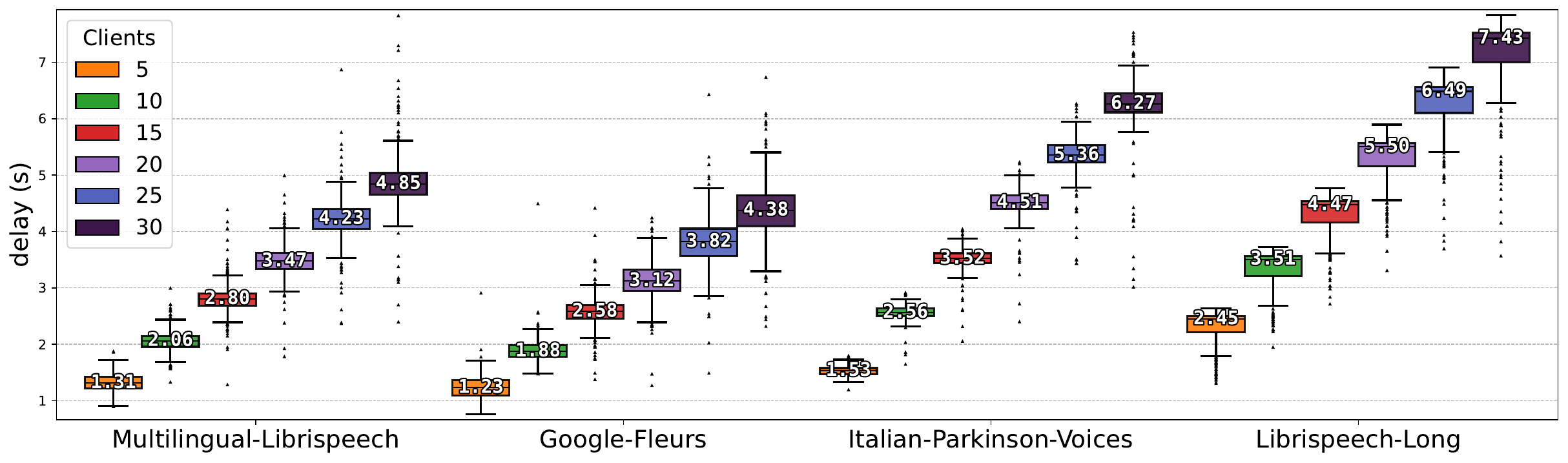}
				\caption{Chunk duration of 0.5s}%
				\label{fig:evaluation-delay0.5}
			\end{subfigure}
			\ \vspace*{0.2cm}
			\begin{subfigure}[b]{\textwidth}
				\centering
				\captionsetup{skip=1pt}
				\includegraphics[width=\textwidth]{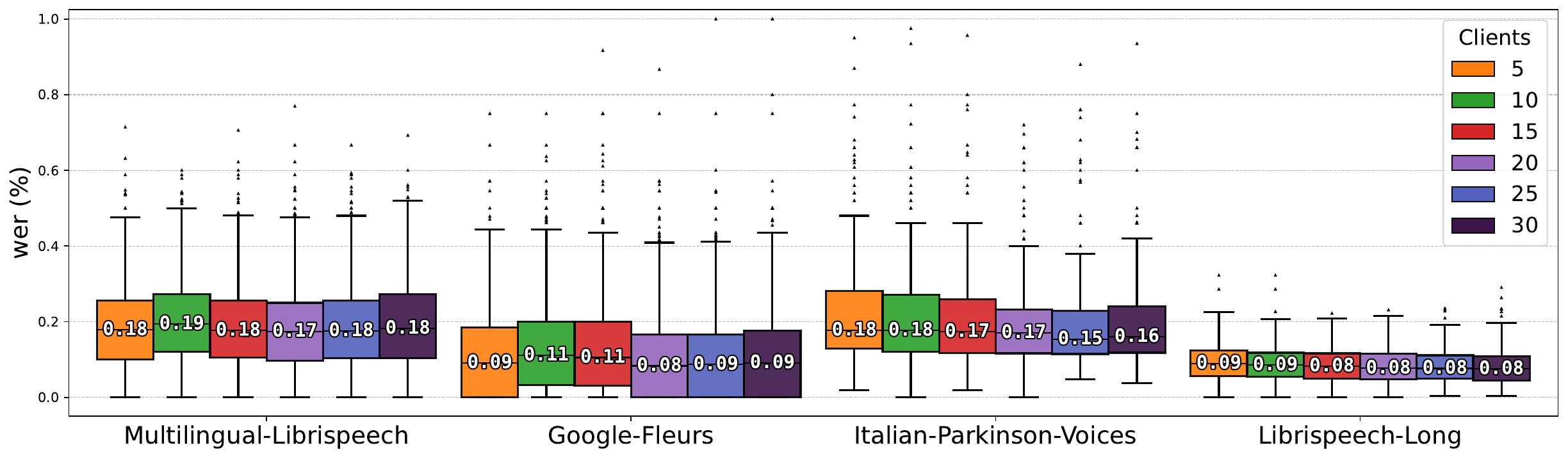}
				\caption{Word Error Rate with chunk duration of 1.0s}%
				\label{fig:evaluation-wer1.0}
			\end{subfigure}
			\ \vspace*{0.2cm}
			\begin{subfigure}[b]{\textwidth}
				\centering
				\captionsetup{skip=1pt}
				\includegraphics[width=\textwidth]{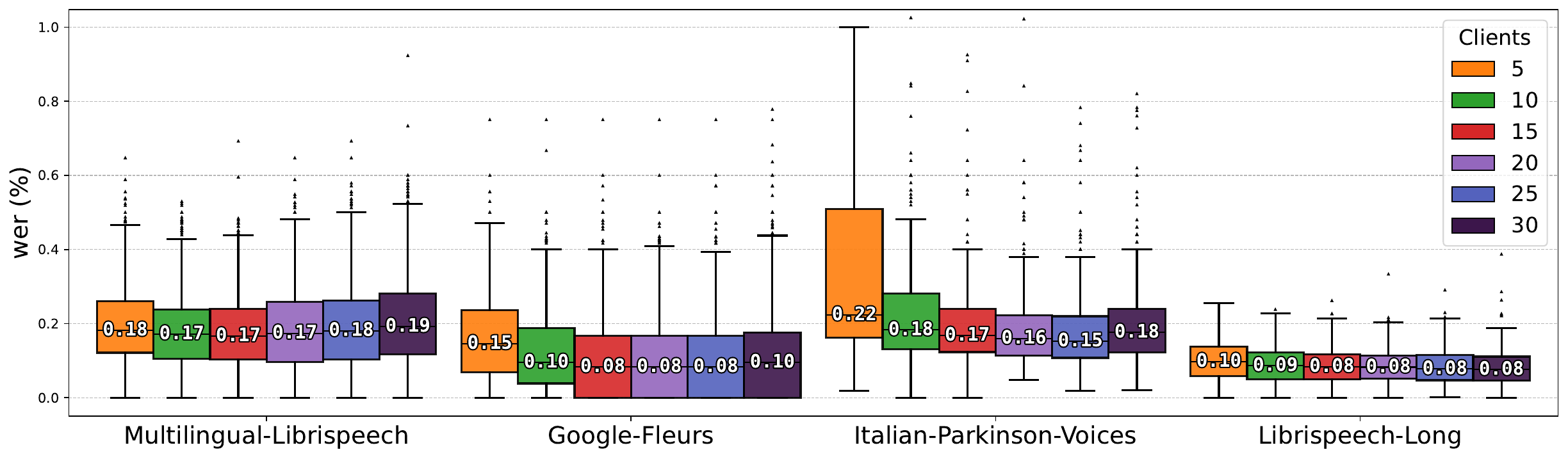}
				\caption{Word Error Rate with chunk duration of 0.5s}%
				\label{fig:evaluation-wer0.5}
			\end{subfigure}
		\end{minipage}
	\end{adjustbox}
	\captionsetup{skip=1pt}
	\caption{Evaluation results with different chunk durations using box plots.}%
	\label{fig:evaluation-all}
\end{figure*}

\begin{figure*}[thbp]
	\centering
	\captionsetup{skip=1pt}

	\begin{subfigure}[t]{0.49\textwidth}
		\centering
		\captionsetup{skip=1pt}
		\includegraphics[width=\linewidth]{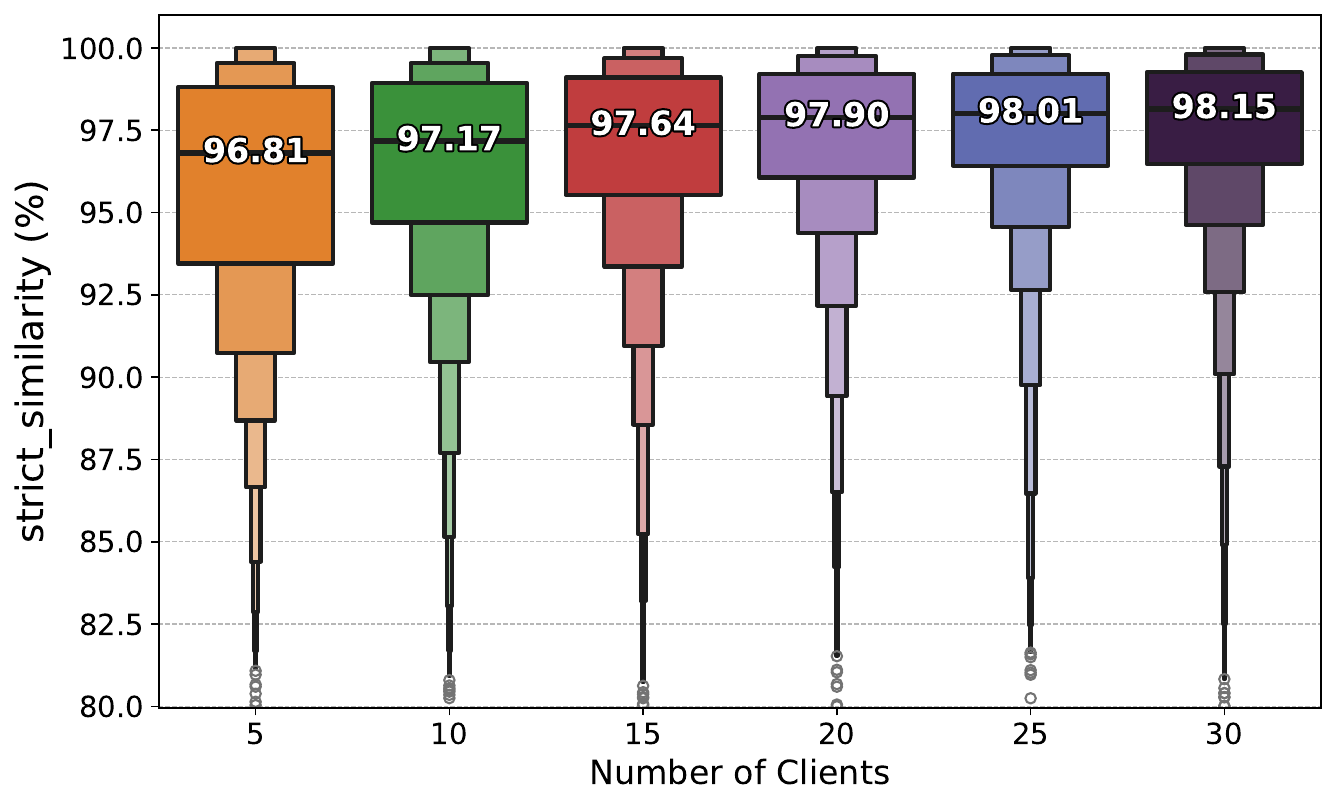}
		\caption{Strict similarity}%
		\label{fig:strict-similarity}
	\end{subfigure}
	\hfill
	\vspace{0.2cm}
	\begin{subfigure}[t]{0.49\textwidth}
		\centering
		\captionsetup{skip=1pt}
		\includegraphics[width=\linewidth]{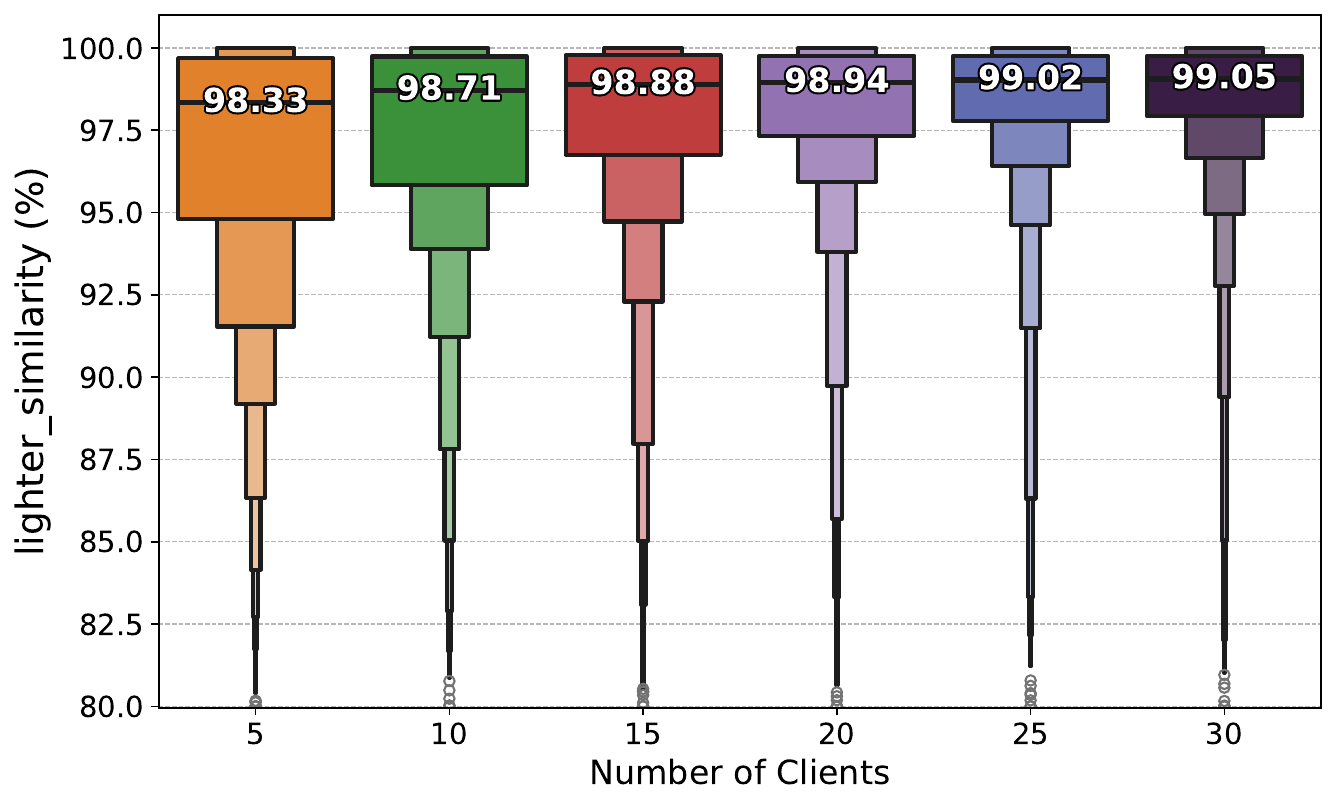}
		\caption{Lighter similarity}%
		\label{fig:lighter-similarity}
	\end{subfigure}

	\caption{The similarity score distribution per client across all datasets using boxen plots.}%
	\label{fig:similarity-distribution}
\end{figure*}

\subsection{Experiments and Results}
\textbf{Delay Distribution Plots.}\quad
Figures~\ref{fig:evaluation-delay1.0} and~\ref{fig:evaluation-delay0.5} show the delay distribution by language under different client loads, using chunk durations of 1.0\,s and 0.5\,s, respectively.
Each box plot group is labeled by dataset on the x-axis, with delay (in seconds) on the y-axis.
While delays generally increase linearly with more concurrent clients, the plots are useful for identifying the threshold where performance starts to degrade.
This insight helps determine when to switch to offline transcription or scale \ourtool{SWIM} by allocating additional resources.

The delays are computed as the mean difference between the timestamp of each confirmed transcription segment and its expected delay from the start of the audio stream. In the plots, colors represent different numbers of concurrent clients.
As expected, both plots show an approximately linear increase in delay as the number of clients grows, due to increased contention for model inference.
Interestingly, with fewer clients, shorter chunks (0.5\,s) yield lower delays by enabling more frequent updates and faster reaction times. However, as concurrency increases, the 1.0\,s chunk configuration begins to outperform the 0.5\,s one, likely due to reduced overhead from less frequent processing.

These results highlight the importance of tuning the chunk duration parameter for optimal performance. Shorter chunks (0.5\,s) are more responsive in low-load scenarios (up to 15 clients), reducing delay by approximately 11\% compared to 1.0\,s chunks. In contrast, for higher loads (more than 15 clients), 1.0\,s chunks yield about 8\% lower delay, due to reduced processing overhead.
Overall, this behavior allows \ourtool{SWIM} to dynamically adapt to workload conditions, balancing throughput and latency as the number of clients changes.\smallskip

\noindent\textbf{WER Distribution Plots.}\quad
Figures~\ref{fig:evaluation-wer1.0} and~\ref{fig:evaluation-wer0.5} present the WER distribution per language for different numbers of concurrent clients, using chunk durations of 1.0\,s and 0.5\,s, respectively.
Each group of box plots is labeled with the corresponding dataset on the x-axis, while the y-axis shows the word error rate.
As expected, the WER remains relatively stable across varying concurrency levels, with only minor fluctuations. This consistency suggests that transcription quality remains largely unaffected by the number of parallel audio streams.
The WER is computed as:\[\text{WER} = \frac{S + D + I}{N}\] where \(S\) is the number of substitutions, \(D\) the number of deletions, \(I\) the number of insertions, and \(N\) the total number of words in the reference transcription.
As expected, WER remains relatively stable across different levels of concurrency, with only minor fluctuations. This behavior suggests that system accuracy is more influenced by architectural choices---particularly the use of a shared buffer---than by the number of concurrent requests.
Notably, \ourtool{SWIM} consistently maintains low WER values in multilingual settings, highlighting one of its key strengths. While chunk duration has some effect on WER, the differences are marginal.

The most notable deviation occurs in the \textit{Italian Parkinson's Voice} dataset, where the WER is higher and more variable for 5 concurrent clients. This can be attributed to the presence of samples where speakers articulate isolated words with long pauses rather than continuous speech. In such cases, when processing latency is low, the limited context can lead to punctuation mistakes---such as incorrect or missing sentence boundaries---which in turn affect casing. These errors disproportionately inflate the WER metric.

Although WER is the \textit{de facto} standard for evaluating ASR accuracy, it does not always reliably capture performance in real-time scenarios. In particular, it penalizes formatting issues---such as punctuation and casing---even when the transcription remains semantically accurate.

Nevertheless, \ourtool{SWIM} maintains low WER across datasets and languages, demonstrating its effectiveness with multilingual inputs and high transcription accuracy.\smallskip

\noindent\textbf{Similarity Distribution Plots.}\quad
Figures~\ref{fig:strict-similarity} and~\ref{fig:lighter-similarity} display the similarity score distribution by number of concurrent clients across all datasets. Scores are shown as percentages and visualized using boxen plots,\footnote{Boxen plots extend traditional box plots by adding additional boxes to better capture the distribution's shape.} which provide a more detailed view of the data distribution.

The expected outcome is that similarity scores---computed using the \texttt{QRatio} metric---will be higher than corresponding WER values, as WER penalizes formatting inconsistencies (e.g., punctuation and casing) that do not impact semantic correctness.
Both similarity scores are computed using the previously introduced \texttt{QRatio} metric.
For each response, we extract the corresponding segment of the reference transcript using its start and end timestamps. We then calculate similarity between this reference segment and the generated response text.
The \textit{strict similarity} score uses the standard \texttt{QRatio}, while the \textit{lighter similarity} score ignores differences in punctuation and casing.
As expected, the strict similarity is slightly lower than the lighter score, but the difference is small, indicating that the system maintains strong semantic accuracy even under stringent comparison criteria.

Interestingly, similar to the trends seen in Figures~\ref{fig:evaluation-wer1.0} and~\ref{fig:evaluation-wer0.5}, similarity scores tend to improve as the number of concurrent clients increases. This is expected because longer processing times allow more audio to accumulate in each dedicated service buffer, giving the model more context when inferring each clip within the shared buffer.

Finally, we observe that both similarity scores remain relatively stable across datasets, with only minor variations. As shown in Fig.~\ref{fig:similarity-distribution}, the median similarity scores exceed \(96\%\) for the strict metric and \(98\%\) for the lighter one, regardless of the number of concurrent clients. This indicates that \ourtool{SWIM} consistently maintains high semantic accuracy across diverse datasets and languages—a crucial requirement for real-time ASR systems.\smallskip

\begin{figure*}[thbp]
	\centering
	\captionsetup{skip=1pt}

	\begin{subfigure}[t]{0.35\textwidth}
		\centering
		\captionsetup{skip=1pt}
		\includegraphics[width=\linewidth]{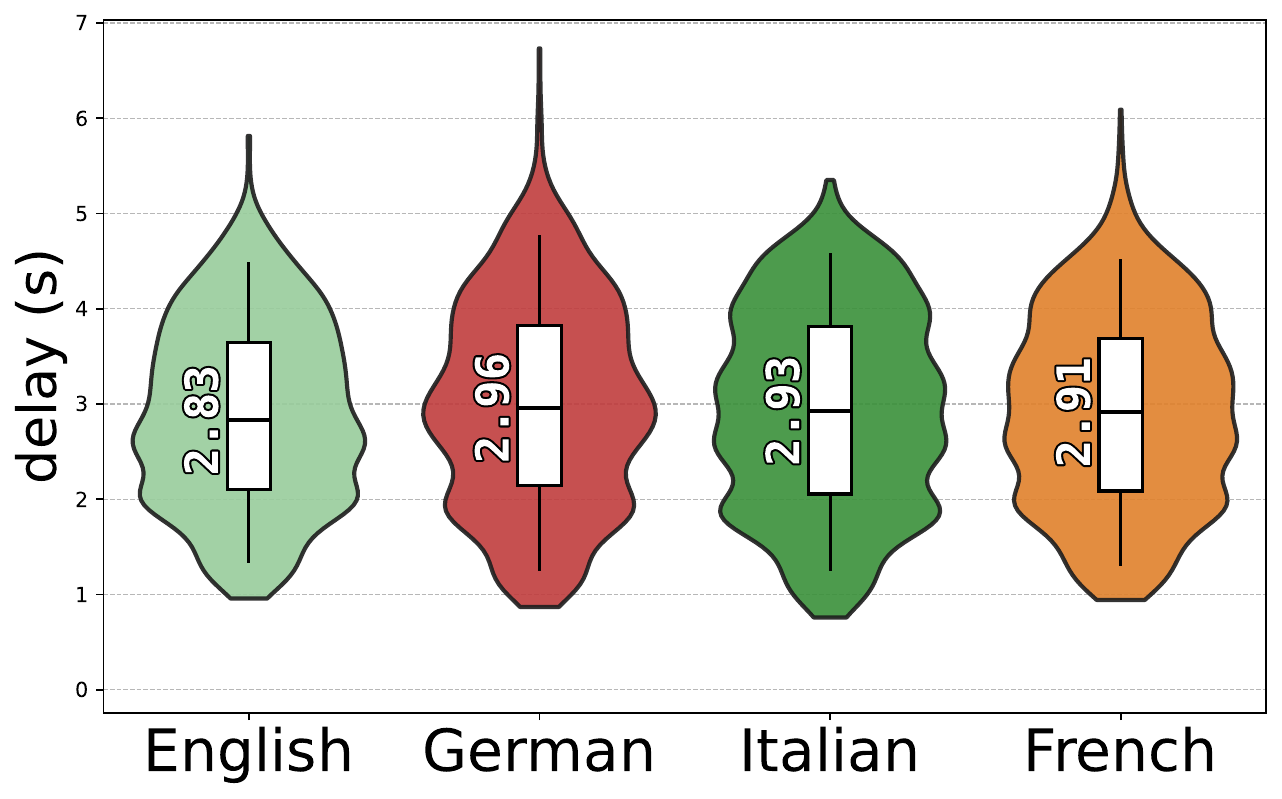}
		\caption{Google Fleurs dataset}%
		\label{fig:delay-gogfleurs-per-language}
	\end{subfigure}
	\hfill
	\vspace{0.2cm}
	\begin{subfigure}[t]{0.59\textwidth}
		\centering
		\captionsetup{skip=1pt}
		\includegraphics[width=\linewidth]{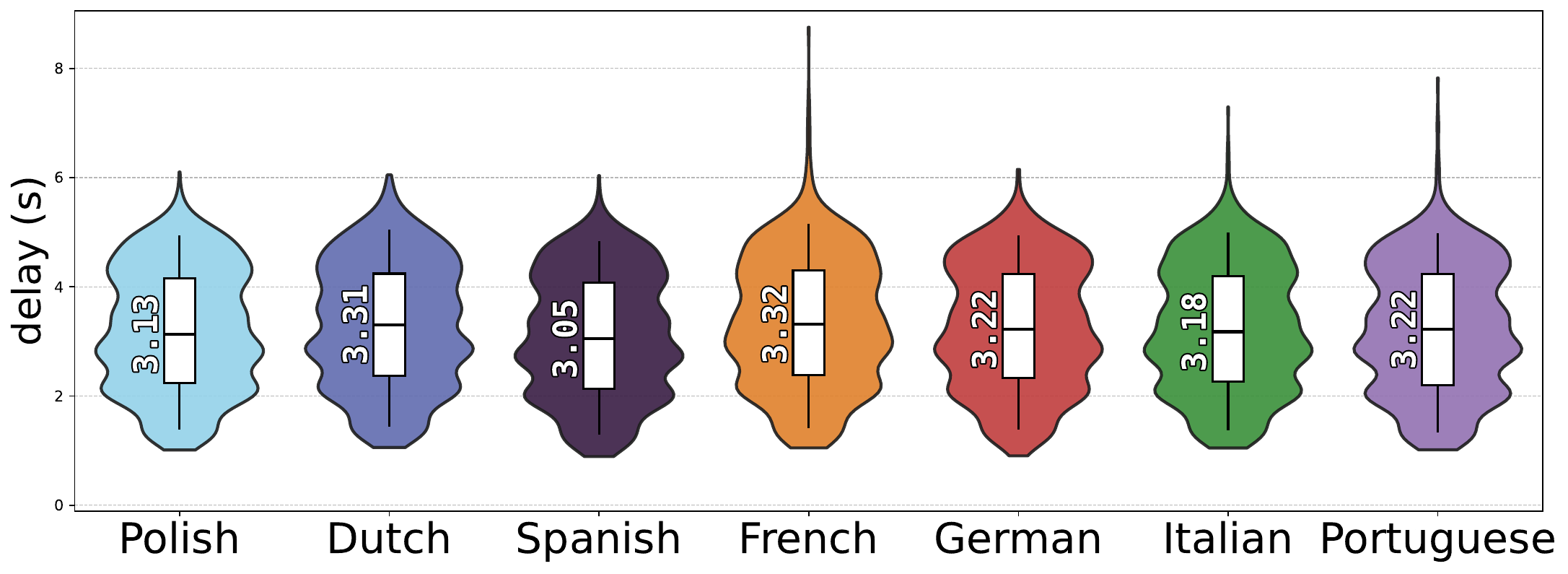}
		\caption{Multilingual LibriSpeech dataset}%
		\label{fig:delay-multilibri-per-language}
	\end{subfigure}

	\caption{Delay per language across the Multilingual LibriSpeech dataset and Google Fleurs dataset using violin plots.}%
	\label{fig:delay-per-language}
\end{figure*}

\begin{figure*}[thbp]
	\centering
	\captionsetup{skip=1pt}

	\begin{subfigure}[t]{0.35\textwidth}
		\centering
		\captionsetup{skip=1pt}
		\includegraphics[width=\linewidth]{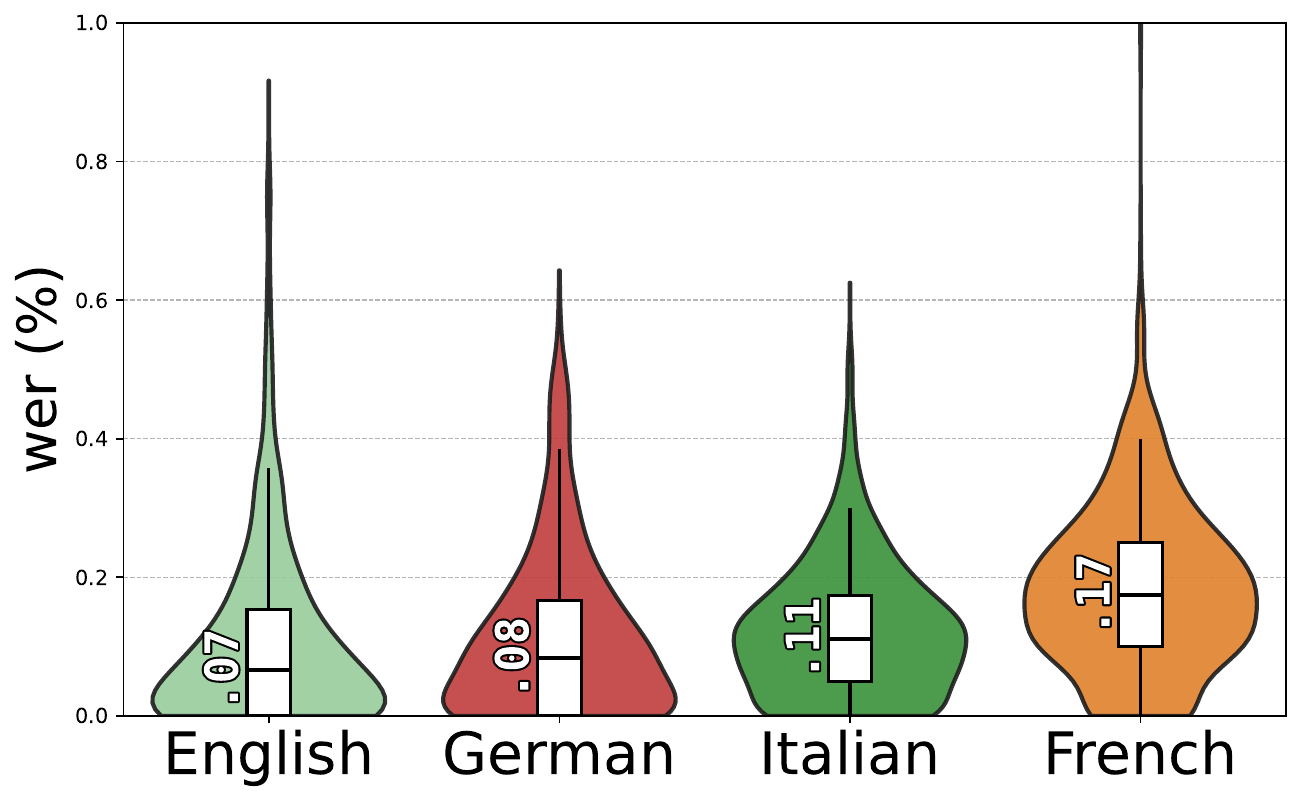}
		\caption{Google Fleurs dataset}%
		\label{fig:wer-gogfleurs-per-language}
	\end{subfigure}
	\hfill
	\vspace{0.2cm}
	\begin{subfigure}[t]{0.59\textwidth}
		\centering
		\captionsetup{skip=1pt}
		\includegraphics[width=\linewidth]{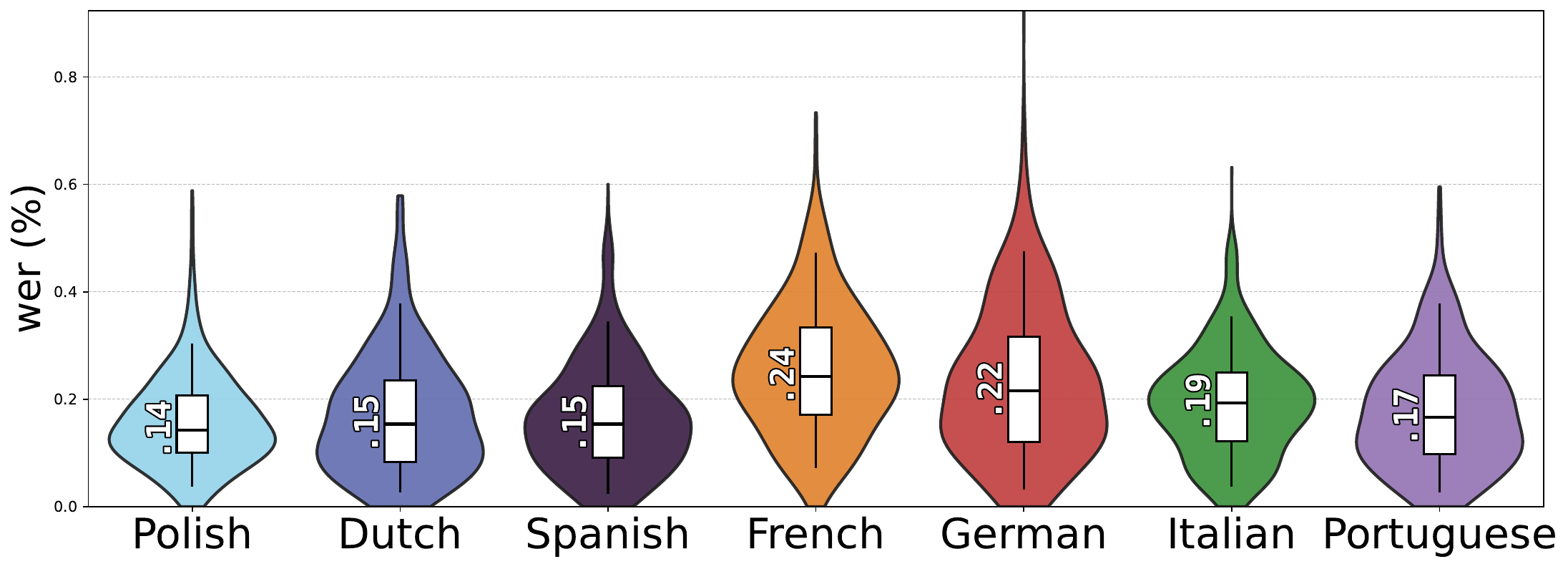}
		\caption{Multilingual LibriSpeech dataset}%
		\label{fig:wer-multilibri-per-language}
	\end{subfigure}

	\caption{WER per language across the Multilingual LibriSpeech dataset and Google Fleurs dataset using violin plots.}%
	\label{fig:wer-per-language}
\end{figure*}

\noindent\textbf{Linguistic Delay Distribution Plots.}\quad
Figure~\ref{fig:delay-gogfleurs-per-language} shows the delay distribution per language for the \textit{Google Fleurs} dataset, while Figure~\ref{fig:delay-multilibri-per-language} presents the same for the \textit{Multilingual LibriSpeech} dataset. Both use violin plots, which combine box plots with kernel density estimation to visualize delay distributions.
Thanks to \ourtool{SWIM}'s client synchronization mechanism, delays are expected to remain consistent across languages, as they are primarily influenced by model processing time.

The most notable observation is that the distributions are similar across languages, with only minor variations in the median delay.
Moreover, the relatively narrow spread of the distributions denotes stable and predictable latency behavior, which is crucial for real-time applications. These results highlight \ourtool{SWIM} generalizes effectively across diverse linguistic contexts without sacrificing responsiveness, underscoring its potential utility in multilingual speech processing environments.\smallskip

\noindent\textbf{Linguistic WER Distribution Plots.}\quad
Figures~\ref{fig:wer-gogfleurs-per-language} and~\ref{fig:wer-multilibri-per-language} show WER distribution across languages in the same settings.

As expected, WER is slightly influenced by the linguistic characteristics of each language. English, for instance, typically yields the lowest WER, while languages with diacritics or complex orthographic rules may show higher error rates.

Nonetheless, the plots reveal that WER distributions remain consistent across languages, with only minor differences in median values. The small variations can be attributed to distribution tails and outliers. This indicates that languages rich in diacritics or with complex orthographies do not significantly degrade \ourtool{SWIM}'s transcription accuracy.

Unlike \texttt{whisper-streaming}, which already demonstrates high accuracy, \ourtool{SWIM} maintains comparable WER across languages---including those with complex phonetic structures---even as the number of concurrent clients increases.\smallskip

\noindent\textbf{Hypothesis vs. Confirmation Delay Plot.}\quad
Figure~\ref{fig:hypothesis-delay} shows the mean delays---visualized as a bar plot---for both hypothesis and confirmation responses on the \textit{Multilingual LibriSpeech} dataset, using a fixed chunk duration of 1.0\,s. The purpose of this plot is to illustrate that, beyond confirmed responses---which represent finalized, immutable transcriptions---hypothesis responses can provide preliminary transcriptions earlier in the pipeline.
Thanks to the local agreement mechanism, hypothesis responses serve as early approximations of the final confirmed outputs. These intermediate results are often sufficiently accurate to be used for immediate feedback to users.
This distinction is especially valuable in real-time applications where low-latency feedback is critical, such as live captioning or when supplying incremental input to \textit{large language models} in interactive agent systems.

\subsection{Concluding Remarks}
\texttt{Whisper-Streaming}~\cite{Machacek23}, in an English-only setting with a single fixed client, achieves a WER of approximately \(8.2\%\) and an average transcription delay of about \(3.4\) seconds.
In comparison, \ourtool{SWIM} scales to \(5\) concurrent clients while maintaining a similar WER (see Tables~\ref{fig:evaluation-wer1.0} and~\ref{fig:evaluation-wer0.5} on the \textit{LibriSpeech-Long dataset}) and reducing average delay to roughly \(2.4\) seconds.
Even at \(15\) concurrent clients, \ourtool{SWIM} sustains stable latency and accuracy comparable to \texttt{Whisper-Streaming}.
Furthermore, \ourtool{SWIM} maintains low delay and high transcription accuracy—measured by WER and similarity scores—across diverse datasets and multilingual scenarios with up to \(20\) concurrent clients.
These results demonstrate \ourtool{SWIM}'s effective scalability, significantly increasing throughput without sacrificing transcription quality.

%% file: sects/threats.tex
\section{Threats to Validity}\label{sect:threats-to-validity}
In our discussion, we follow \citet{Wohlin12}'s taxonomy.

\subsection{Construct Validity}\label{sect:construct-validity}
\noindent\textbf{Transcription Quality Metrics.}\quad Our evaluation also relies on standard ASR metrics such as WER to assess transcription quality. However, these metrics may not fully reflect the nuances of real-time multilingual transcription, especially in cases involving code-switching, domain-specific vocabulary, or varied audio conditions across clients.\smallskip

\noindent\textbf{Real-Time Performance Metrics.}\quad
We use latency and throughput to evaluate real-time performance. However, the definition of ``real-time'' varies by application. Our processing thresholds may not meet the stricter requirements of scenarios like live broadcasting or emergency response systems, where sub-second latency is critical.

\noindent\textbf{Scalability Assessment.}\quad
Our evaluation measures scalability by the number of concurrent clients the system can support. This may not fully reflect real-world conditions, where clients vary in audio quality, language, and network stability.\smallskip

\noindent\textbf{Mitigation Strategies.}\quad
To address these threats to construct validity, we adopted several strategies: \begin{inparaenum}
   \item for transcription quality, we complement standard WER metrics with our QRatio-based local agreement mechanism and a fallback confirmation strategy within the adaptive hypothesis buffer. This enables more nuanced handling of multilingual and code-switched input, especially in noisy or ambiguous contexts;
   \item for real-time performance, our latency threshold applies only to confirmed segments, while unconfirmed portions are updated dynamically using word-level indicators. This better reflects practical expectations for streaming ASR systems;
 nd  \item for scalability, \ourtool{SWIM} was evaluated on four datasets (including two multilingual) recorded in both studio and real-world conditions, covering diverse speakers and varying audio quality. This helps ensure that our results generalize to real-world deployments rather than idealized settings.
\end{inparaenum}

\begin{figure}[t]
   \centering
   \includegraphics[width=\linewidth]{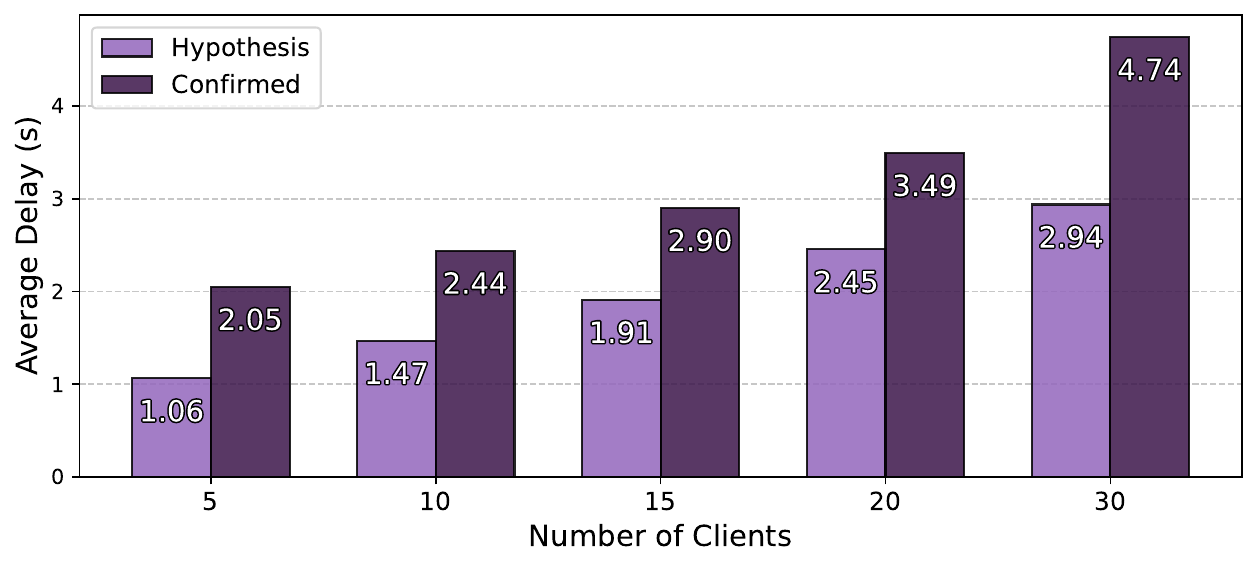}
   \captionsetup{skip=1pt}
   \caption{Hypothesis delay vs Confirmation delay with fixed chunk duration of 1.0s on Multilingual LibriSpeech dataset.}%
   \label{fig:hypothesis-delay}
\end{figure}

\subsection{Internal Validity}\label{sect:internal-validity}
\textbf{Whisper Model Limitations.}\quad
The underlying \texttt{Whisper} model has known limitations when handling silent audio segments, occasionally producing hallucinated or repeated phrases. These artifacts can inflate error rates and may not accurately reflect the performance of the \ourtool{SWIM} system architecture.\smallskip

\noindent\textbf{Audio Quality Variations.}\quad
Inconsistent audio quality across test samples may affect our evaluation. Background noise, microphone fidelity, speaker accents, and recording conditions can introduce biases, favoring certain client configurations or audio types.\smallskip

\noindent\textbf{Mitigation Strategies.}\quad
To address these threats to internal validity, we adopted several strategies:
\begin{inparaenum}
   \item to mitigate limitations of the \texttt{Whisper} model, we support optional voice activity detection as a preprocessing step. This filters out non-speech segments, reducing hallucinations caused by silence; and
   \item to handle audio quality variations, \ourtool{SWIM} includes a fallback mechanism that activates after repeated local agreement failures, ensuring reliable output even under challenging acoustic conditions.
\end{inparaenum}

\subsection{External Validity}\label{sect:external-validity}
\textbf{Dataset Limitations.}\quad
Our evaluation relies on selected datasets that may not capture the full diversity of real-world audio streams. If these datasets predominantly feature clean, studio-quality recordings or limited language combinations, the results may not generalize to noisy environments, telephone audio, or underrepresented languages in \texttt{Whisper}'s training data.\smallskip

\noindent\textbf{Client Configuration Constraints.}\quad
Our experiments assume clients use uniform audio chunk durations (\(\leq 1\) second) and fixed sampling rates (\(16\)kHz). However, real-world deployments may involve more diverse configurations, including varying chunk lengths and audio formats, which could impact system performance.\smallskip

\noindent\textbf{Deployment Environment.}\quad
Our evaluation takes place in controlled server environments that may not capture the constraints of real-world deployments, such as limited computational resources, fluctuating network conditions, and varying system loads.\smallskip

\noindent\textbf{Mitigation Strategies.}\quad
To address threats to external validity, we implemented several strategies:
\begin{inparaenum}
   \item to overcome dataset limitations, our evaluation covers diverse audio conditions---including noisy environments, telephony audio, and multiple languages—beyond clean, studio-quality recordings to improve generalizability;
   \item concerning client configuration constraints, \ourtool{SWIM} supports variable chunk sizes, with some settings yielding better performance. Although a \(16\)kHz sampling rate is currently required, we are extending support for automatic upsampling and downsampling to increase compatibility with different input formats; and
   \item while testing occurred in controlled environments, \ourtool{SWIM} is actively deployed in production by \textbf{REDACTED}, serving both internal and external clients, confirming its real-world effectiveness.
\end{inparaenum}

\subsection{Conclusion Validity}\label{sect:conclusion-validity}
\textbf{Baseline Comparison Limitations.}\quad
Comparisons with existing systems such as \texttt{Whisper-Streaming} and \texttt{Whispy} may be affected by differences in implementation, evaluation settings, or optimization goals, which can make direct performance comparisons potentially misleading.\smallskip

\noindent\textbf{Statistical Analysis Limitations.}\quad
Our analysis is limited by sample size, which may affect the reliability of our conclusions. Small samples can yield unstable performance estimates and may not fully capture true variability in system behavior.\smallskip

\noindent\textbf{Mitigation Strategies.}\quad
To address threats to conclusion validity, we adopted several strategies:
\begin{inparaenum}
   \item although implementation differences can impact baseline comparisons, our primary contribution is a novel architectural approach to multi-client ASR, which can be adopted or adapted by other systems regardless of their implementation. This highlights methodological innovation over direct performance advantage; and
   \item to strengthen statistical validity, we evaluated \ourtool{SWIM} on datasets with long audio samples (up to 5 minutes) and hundreds of entries, tested across varying numbers of concurrent clients. This reduces small-sample effects and offers a more comprehensive assessment of system behavior under diverse load conditions.
\end{inparaenum}

%% file: sects/conclusion.tex
\section{Conclusion}\label{sec:conclusion}
We present \ourtool{SWIM}, a novel multi-client real-time ASR system that enables a single \texttt{Whisper} model instance to process multiple concurrent multilingual streams with low latency and high accuracy. Unlike multi-instance setups, \ourtool{SWIM} achieves model-level parallelization through a shared buffer merging strategy, optimizing resource utilization while preserving transcription quality. Key innovations include an adaptive hypothesis buffer with \texttt{QRatio}-based agreement and a fallback confirmation mechanism to manage linguistic variability.\ \ourtool{SWIM} outperforms existing single-client methods and provides scalable, cost-effective ASR suitable for enterprise deployment.